\documentclass[]{pasj01}

\begin{document} 
\Received{}
\Accepted{}

\title{ULTRAVIOLET TO OPTICAL DIFFUSE SKY EMISSION AS SEEN BY THE {\it HUBBLE SPACE TELESCOPE} FAINT OBJECT SPECTROGRAPH}

\author{K. \textsc{Kawara}\altaffilmark{1}}%
\altaffiltext{1}{Institute of Astronomy, University of Tokyo, 2-21-1, Osawa, Mitaka, Tokyo 181-0015, Japan}

\author{Y. \textsc{Matsuoka}\altaffilmark{2}}
\altaffiltext{2}{National Astronomical Observatory of Japan, 2-21-1 Osawa, Mitaka, Tokyo 181-8588, Japan}

\author{K. \textsc{Sano}\altaffilmark{3,4,*}}
\altaffiltext{3}{Department of Astronomy, Graduate School of Science, The University of Tokyo, Hongo 7-3-1, Bunkyo-ku, Tokyo 113-0033, Japan}
\altaffiltext{4}{Institute of Space and Astronautical Science, Japan Aerospace Exploration Agency, 3-1-1 Yoshinodai, Chuo-ku, Sagamihara, Kanagawa 252-5210, Japan}
\email{sano@ir.isas.jaxa.jp}

\author{T. \textsc{Brandt}\altaffilmark{5}}
\altaffiltext{5}{Institute for Advanced Study, Einstein Dr., Princeton, NJ, USA}

\author{H. \textsc{Sameshima}\altaffilmark{6}}
\altaffiltext{6}{Koyama Astronomical Observatory, Kyoto Sangyo University,  Motoyama, Kamigamo, Kita-ku, Kyoto, 603-8555, Japan}

\author{K. \textsc{Tsumura}\altaffilmark{7}}
\altaffiltext{7}{Frontier Research Institute for Interdisciplinary Science, Tohoku University, Sendai 980-8578, Japan}

\author{S. \textsc{Oyabu}\altaffilmark{8}}
\altaffiltext{8}{Nagoya University, Furo-cho, Chikusa-ku, Nagoya 464-8601, Japan}

\author{N. \textsc{Ienaka}\altaffilmark{1}}



\maketitle

\begin{abstract}
We present an analysis of the blank sky spectra observed with the Faint Object Spectrograph on board the {\it Hubble Space Telescope}. 
We study the diffuse sky emission from ultraviolet to optical wavelengths, which is composed of the zodiacal light (ZL), diffuse Galactic light (DGL), and residual emission.
The observations were performed toward $54$ fields distributed widely over the sky, with the spectral coverage from $0.2$ to $0.7\,\rm{\mu m}$.
In order to avoid contaminating light from the earthshine, we use the data collected only in orbital nighttime. 
The observed intensity is decomposed into the ZL, DGL, and residual emission, in eight photometric bands spanning our spectral coverage. 
We found that the derived ZL reflectance spectrum is flat in the optical, 
which indicates major contribution of C-type asteroids to the interplanetary dust (IPD).
In addition, the ZL reflectance spectrum has an absorption feature at $\sim0.3\,\rm{\mu m}$.
The shape of the DGL spectrum is consistent with those found in earlier measurements and model predictions. 
While the residual emission contains a contribution from the extragalactic background light, we found that the spectral shape of the residual looks similar to the ZL spectrum.
Moreover, its optical intensity is much higher than that measured from beyond the IPD cloud by {\it Pioneer10/11}, and also than that of the integrated galaxy light.
These findings may indicate the presence of an isotropic ZL component, which is missed in the conventional ZL models. 
\end{abstract}

\section{Dedication}

This work was suggested and initiated by Kimiaki Kawara, with the original aim of measuring the extragalactic background light in the ultraviolet to optical wavelengths.
For several years, he made concentrated efforts 
with great patience and has written up a draft of this paper.
Unfortunately, due to his failing health, he was not able to see its publication 
before he passed away in January 2015. 
This paper has subsequently been completed by his colleagues, and is here dedicated to his memory.

\section{Introduction}


Measurements of the diffuse sky emission are important for probing various astrophysical phenomena, such as interstellar dust emission and extragalactic 
background light (EBL), which complement observations of discrete sources (stars, galaxies, etc.) to shape our understanding of the universe.
From ultraviolet (UV) to optical wavelengths, the diffuse sky emission consists of the airglow, the zodiacal light (ZL), the diffuse Galactic light (DGL), and the residual emission including the EBL.

\subsection{Zodiacal Light}

The ZL is the brightest emission component of the diffuse sky brightness, when observations are made from the space.
The ZL consists of the sunlight scattered by the interplanetary dust (IPD) and thermal emission from the IPD.
From the UV to optical wavelengths, the ZL brightness is dominated by the scattered sunlight.

The IPD is expected to fall into the sun by the Poynting-Robertson drag and also leave the solar system by the radiation pressure, in a time scale of $10^3$--$10^7$ years, 
which is much shorter than the age of the solar system (Mann et al. 2006). 
Therefore, the IPD particles 
should be supplied continuously by asteroids or comets, though the contribution of each component to the IPD is unclear.
Comparison of the reflectance spectrum of the ZL with those of asteroids and comets is a useful way to identify the IPD supplier(s).
By combining the ZL reflectance measurements with the Low Resolution Spectrometer (LRS) on board Cosmic Infrared Background Experiment (CIBER) and those with {\it Infrared Telescope in Space} ({\it IRTS}), Tsumura et al. (2010) suggested that the spectral shape of the ZL reflectance is similar to that of a S-type asteroid (Binzel et al. 2001) from the red-optical to near-infrared (IR) wavelengths ($\sim0.8$--$2.5\,\rm{\mu m}$).
On the other hand, Yang and Ishiguro (2015) reported that the ZL reflectance spectrum in the near-IR is similar to that of comets, which are classified as D-type asteroids (Bus \& Binzel 2002).

Measurements of the ZL reflectance spectrum in the optical are a key to understanding the origin of the IPD, since
different types of asteroids have significantly different reflectance spectra 
in this wavelength range (Bus \& Binzel 2002).
For example, C-type asteroids have much flatter reflectance spectrum than do S-type asteroids.
However, there has been no optical measurements of the ZL reflectance spectrum published to date. 

\subsection{Diffuse Galactic Light}

The DGL consists of starlight scattered by, and re-radiated as thermal emission from, the interstellar dust grains in the diffuse interstellar medium (ISM).
The scattered component dominates in the UV to optical wavelength range.
DGL measurements are useful to constrain the properties of the interstellar dust, such as the size distribution and grain albedo.
Interstellar $100\,\rm{\mu m}$ dust emission has been used as a tracer of the DGL emission, since the intensities of these two emissions 
are expected to correlate linearly 
in the optically thin limit (Brandt \& Draine 2012).

Recent analyses have detected the DGL from the optical to near-IR wavelengths, using the data obtained with {\it Pioneer10/11} (Matsuoka et al. 2011), CIBER (Arai et al. 2015), and the Diffuse Infrared Background Experiment (DIRBE) on board {\it Cosmic Background Explorer} ( {\it COBE}; Sano et al. 2015).
These results are marginally consistent with the DGL model spectra presented by Brandt \& Draine (2012), which are based on the interstellar dust models of Weingartner \& Draine (2001) and Zubko et al. (2004).

\subsection{Residual Emission}

The residual emission, which is obtained by subtracting all the foreground components (the ZL and DGL in the case of the space observations) 
from the observed diffuse sky emission, 
contains the EBL.
The EBL is the cumulative emission from known extragalactic sources such as galaxies, intergalactic matter, protogalaxies, and various pregalactic objects.
It may also contain radiation originating from decays of elementary particles, such as sterile neutrinos (Mapelli \& Ferrara 2005).
The EBL is thus a fundamental quantity to constrain the energy emitted from the entire cosmological objects, which is important to understand the evolution of the Universe and galaxies.

The contribution of known galaxies to the EBL can be estimated by integrating the galaxy luminosity function; this is particularly true in the UV to near-IR wavelengths, 
thanks to deep galaxy counts obtained from space and ground-based observations (e.g., Gardner et al. 2000; Madau \& Pozzetti 2000; Totani et al. 2001; Dom\`inguez et al. 2011). 
Diffuse emission measurements have sometimes reported the residual emission much stronger than the integrated galaxy light (IGL); 
such measurements include those by 
Bernstein (2007) using the optical imaging data taken with the Wide Field Planetary Camera 2 (WFPC2) on board {\it Hubble Space Telescope} ({\it HST}), Matsumoto et al. (2005; 2015) using the near-IR spectroscopic data taken with the Near-Infrared Spectrometer (NIRS) on board {\it IRTS}, and Gorjian, Wright, and Chary (2000), Wright and Reese (2000), Wright (2001), Cambr\'esy et al. (2001), Levenson, Wright, and Johnson (2007), and Sano et al. (2015; 2016a) using the near-IR imaging maps taken with {\it COBE}/DIRBE. 
Such a strong residual emission is, if interpreted as the EBL, in conflict with the EBL upper limit obtained by means of the intergalactic attenuation of $\gamma$-rays photons
from distant blazers (e.g., Aharonian et al. 2006; Albert et al. 2008; Abramowski et al. 2013). 
An up-to-date result of the galaxy number counts from the far-UV to far-IR wavelengths has been presented by Driver et al. (2016). 
The resultant IGL brightness is consistent with earlier measurements, and is much fainter than the residual diffuse emission obtained by most of the direct measurements
in the optical to near-IR.

As pointed out by Dwek, Arendt, and Krennrich (2005), the spectrum of the near-IR residual emission derived from the {\it IRTS}/NIRS (Matsumoto et al. 2005) measurements 
is similar to that of the ZL.
This may imply a possibility that the residual emission contains an additional ZL component, which is missed in the commonly used DIRBE ZL model (Kelsall et al. 1998).
The DIRBE ZL model was developed by fitting only the time variation of the sky brightness observed with {\it COBE}/DIRBE, in order to determine the physical parameters of the IPD structures.
Therefore, it cannot uniquely determine the true ZL signal, because an arbitrary amount of an isotropic ZL component could be added (Hauser et al. 1998).

Strong residual emission has also been reported in the optical wavelength, by Bernstein (2007).
However, since the
results were obtained in only three wide photometric bands at $0.300$, $0.555$, and $0.814\,\rm{\mu m}$,
it is difficult to study the detailed spectral shape of the residual emission.
If we observe an overall similarity in the spectral shapes of the residual and ZL emissions through the optical to near-IR wavelengths, then it would further strengthen 
the hypothesis that a missed ZL component contributes the residual emission.

Another constraint on the optical EBL was presented by Matsuoka et al. (2011), who analyzed the data taken with NASA's {\it Pioneer 10/11} spacecrafts.
{\it Pioneer 10/11} are the first spacecrafts to travel beyond the Asteroid Belt and explore the outer solar system. 
Matsuoka et al. (2011) made use of the optical imaging data at $0.44$ and $0.64\,\rm{\mu m}$, observed at a heliocentric distance $\gtrsim 3.26\,{\rm AU}$.
They performed accurate starlight subtraction and decomposed the observed intensity into the DGL and the residual emission. 
Because the ZL is very faint and  below the detection limit of the instruments beyond $3.26\,{\rm AU}$ (Hanner et al 1974),
the residual emission is free from ZL contamination and can be regarded as the EBL.  
The derived EBL intensity is much lower than the estimates of Bernstein (2007), and are comparable to the IGL intensity. 

\subsection{Outline of the Present Paper}

This paper presents a new analysis of the ZL, DGL, and residual emission from the UV to optical wavelengths (0.2 - 0.7 $\mu$m), using the blank sky spectra obtained 
with the Faint Object Spectrograph (FOS) on board {\it HST}.
We successfully determine the spectral shapes of the individual components, and discuss their implications.
In particular, we derive the ZL reflectance spectrum in the above wavelength range for the first time, and also demonstrate that the spectral shape of the residual emission 
is very similar to that of the known ZL.

This paper is organized as follows: Section 3 describes the {\it HST}/FOS observations, the archival data, and the impact of the earthshine
on the observed sky brightness. 
In Section 4, we 
decompose the observed intensity into the ZL, DGL, and residual emission, using models of the individual components.
This calculation is performed in eight photometric bands, whose central wavelengths range from $0.23$ to $0.65\,\rm{\mu m}$.
After the derived ZL and DGL are discussed in Section 5 and 6, respectively, 
we show the spectrum of the residual emission in comparison with earlier measurements and the ZL spectrum in Section 7.
The summary and conclusion appear in Section 8.

The FOS data are stored in specific intensity $I_{\lambda}$ (${\rm ergs\ cm^{-2}\,s^{-1}\,\AA^{-1}\,arcsec^{-2}}$), while IR and UV data are frequently presented 
in $I_{\nu}$ (${\rm MJy\,sr^{-1}}$) and $N_\lambda$ (${\rm photons\,cm^{-2}\,s^{-1}\,\AA^{-1}\,sr^{-1}}$), respectively.
In this paper, most of the results are presented in $\nu I_{\nu}$ (${\rm nW\,m^{-2}\,sr^{-1}}$).
The conversion formula between these units are:\\
$\nu I_{\nu}({\rm nW\,m^{-2}\,sr^{-1}}) = [3000/\lambda({\rm \mu m})]I_{\nu}({\rm MJy\,sr^{-1}})$,\\
$\nu I_{\nu}({\rm nW\,m^{-2}\,sr^{-1}}) = 0.02\,N_{\lambda}({\rm photons\,cm^{-2}\,s^{-1}\,\AA^{-1}\,sr^{-1}})$, \\
$I_{\nu}({\rm MJy\,sr^{-1}}) = 1.42\times 10^9 \lambda(\rm{\AA})^2 I_{\lambda}({\rm ergs\,cm^{-2}\,s^{-1}\,\AA^{-1}\,arcsec^{-2}})$.
In addition, Ly$\alpha$ and H$\alpha$ emission strengths are expressed in units of Rayleighs (R), where $1{\rm R} = 10^{10}/4\pi\,{\rm photons\,m^{-2}\,s^{-1}\,sr^{-1}}$.

\section{DATA}

\subsection{Observations}

We use the blank sky spectra taken with the FOS on board the {\it HST}.
These data were obtained in Science Verification (SV) observations in $1991$--$1992$, prior to installation of the Corrective Optics Space 
Telescope Axial Replacement (COSTAR; installed in December 1993). 
The relevant proposal information is summarized 
in Table 1. 
Lyons et al. (1992) and Lyons et al. (1993a) used these data to discuss dependencies of the sky background on telescope pointing directions. 
We retrieved the calibrated data from the {\it HST} data archive, which were processed with the Space Telescope Science Data Analysis System (STSDAS) {\it calfos} task.

\begin{table*}
 \renewcommand{\arraystretch}{1.0}
 \caption{FOS SV observations  used in the present analysis}
\begin{center}
  \label{symbols}
  \scalebox{0.9}{
  \begin{tabular}{lcccc}
  \hline
   ProID & Proposal title & $N_{{\rm fld}}$ & $N_{{\rm exp}}$ & $T_{{\rm ksec}}$ \\
   \hline
   SV2965 & FOS-20 SKY BACKGROUND - HIGH GALACTIC LATITUDE & $30$ & $374$ & $47.5$\\
   SV2966 & FOS-21 SKY BACKGROUND - LOW GALACTIC LATITUDE & $8$ & $132$ & $17.0$\\
   SV2967 & FOS-22 SKY BACKGROUND - LOW ECLIPTIC LATITUDE & $16$ & $222$ & $28.5$\\
   Total &  & $54$ & $728$ & $93.0$\\
    \hline
    \end{tabular}
    }
    \end{center}
    \medskip
    
    Note. --- Col 1: Proposal identification number.\\
    Col 2: Proposal title.\\
    Col 3: Total number of observed fields.\\
    Col 4: Total number of exposures.\\
    Col 5: Total exposure time in ksec.
     
 \end{table*}

The FOS was one of the original {\it HST} instruments. 
It had two digicon detectors with independent optical paths, which were photon counting detectors operated by accelerating photoelectrons emitted from a two-dimensional transmissive photocathode onto a linear array of 512 silicon diodes. 
The blue digicon (BLUE) photocathode was sensitive from $1150$ to $5400\,\rm{\AA}$, while the red digicon (RED) photocathode covered the wavelength range from $1620$ to $8500\,\rm{\AA}$. 
The largest entrance aperture of $4_{\cdot}''3\times 4_{\cdot}''3$ was used. 
Since the diode array extended only $1_{\cdot}''43$ in the direction perpendicular to the dispersion, the aperture had an effective collecting area\footnote{The physical dimensions of the individual diodes of the digicons corresponded to spacings $0_{\cdot}''35$ along the dispersion direction and height of $1_{\cdot}''43$ perpendicular to it.} of $4_{\cdot}''3\times 1_{\cdot}''43$. 
In this paper, we analyze spectra taken with the low resolution dispersers, namely, G650L grating and PRISM, combined with the RED digicon. 
The spectral coverages with these disperser configurations are 3540 -- 7075 \AA\ and 1850 -- 8500 \AA, respectively.

%

The blank sky spectra were observed in a standard manner.
They were deflected by a quarter of a diode in the dispersion direction (NXSTEPS = 4) to better sampling and was shifted in the dispersion direction so that five separated diodes contributed to each spectral pixel (OVERSCAN = 5). 
In this way, the 512 diode array produced readout at 2064 spectral positions. 
The ACCUM mode was used, where spectra were read out at specified intervals (typically
a few minutes) and the accumulated sum after each read was stored and recorded in consecutive groups in the standard output data files; each consecutive spectrum was made up of the sum of all previous intervals of data in an ACCUM observation.

The exposures (i.e., integration per readout) were made in the same way for all the relevant observations (see Lyons et al. 1992, for details). 
The telescope stayed still at the pointing direction established prior to the first exposure. 
All exposures were taken in the sequence of G650L and then PRISM at the same pointing direction. 
For each disperser, the first ACCUMlated data file contains 2 exposures totaling 300 seconds (i.e., 150 seconds per exposure), the second contains 5 exposures totaling 600 seconds, and the third contains 7 exposures totaling 900 seconds.

\subsection{Earthshine}

\begin{figure}
\begin{center}
 \includegraphics[scale=0.45]{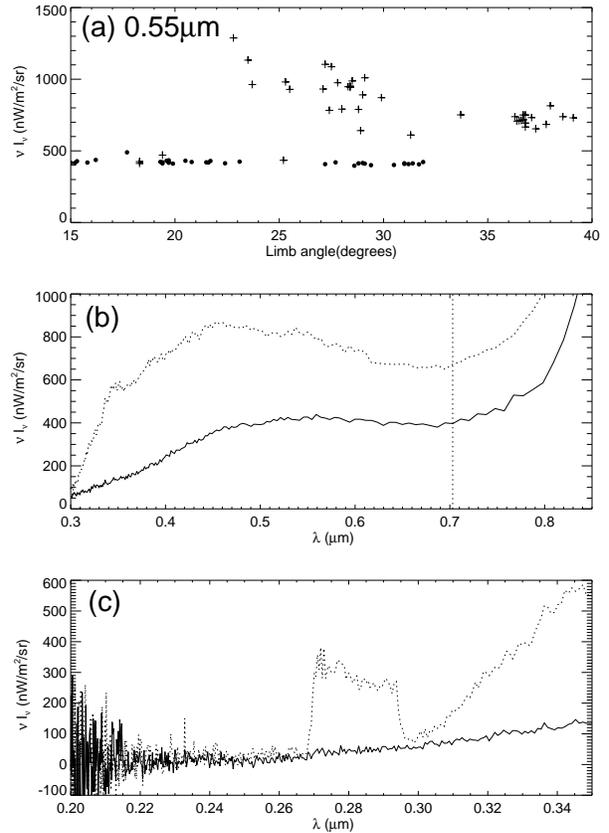} 
\end{center}
 \caption
 {FOS sky data taken at the parallel HDF field with the PRISM/RED configuration. 
Panel (a) plots the sky brightness at $0.55\,\rm{\mu m}$ as a function of limb angle; 
the orbital daytime and twilight/night observations are represented by the crosses and the dots, respectively. 
The following two panels display the mean daytime (the dotted lines) and twilight/nighttime (the solid lines) spectra from $0.3$ to $0.85\,\rm{\mu m}$ (panel b)
and from $0.2$ to $0.35\,\rm{\mu m}$ (panel c). }
\end{figure}

The {\it HST} altitude above the surface of the Earth ranges from $580$ to $630\,\rm{km}$. 
The orbit is inclined at $28_{\cdot}^{\circ}5$ from the equator. 
The telescope completes one orbit in every 96 minutes, passing into the Earth shadow in each orbit, with the time in shadow (orbital night) varying from 28 to 36 minutes. 
Because the equatorial radius of the Earth is approximately $6378\,\rm{km}$, the earth limb is located at approximately $24^{\circ}$ below the horizon.
Data taken during orbital days have significant contamination from earthshine, namely, scattered sunlight in the upper atmosphere and geocoronal emission lines such as Ly$\alpha$ and O I  (e.g., Lyons et al. 1993a; Shaw et al. 1998; Brown et al. 2000).

Here we demonstrate the effect of the earthshine, 
using the high-quality sky data obtained in the parallel mode observations (proposal ID = 6339) of the {\it Hubble} Deep Field (HDF).
These observations were carried out in 1995 December $18$--$28$, after installation of the COSTAR, for engineering purpose.
The pointing center was $(l,b) = (126_{\cdot}^{\circ}1, 54_{\cdot}^{\circ}9)$ in Galactic coordinates and $(\lambda, \beta) = (148_{\cdot}^{\circ}3, 57_{\cdot}^{\circ}2)$ in ecliptic coordinates.
The PRISM and the RED-digicon were used, with the integration times of $46$ and $57\,\rm{ksec}$ for the day and night spectra, respectively. 
The same aperture as in our SV observations was used, but the effective collecting area 
was smaller ($3_{\cdot}''7\times 1_{\cdot}''3$) due to the COSTAR installation.

Figure 1a compares the daytime and twilight/nighttime sky brightness at $0.55\,\rm{\mu m}$, as a function of limb angle, which is the angle between the target field and 
the Earth limb. 
We use the criteria of the daytime fraction (see below) 100 \% for orbital day and 0 -- 50 \% for twilight/nighttime.
Figure 1b compares the daytime and twilight/nighttime spectra (the sharp spectral rise beyond $0.7\,\rm{\mu m}$ is an artifact; see Welsh et al. 1998).
These two panels clearly demonstrate that the daytime spectrum is brighter and bluer than the twilight/nighttime spectrum.  
Figure 1c displays the daytime and twilight/nighttime spectra in the bluest part of our spectral coverage.
An excess emission is seen around $0.28\,\rm{\mu m}$ in the daytime, whose profile is similar to those of geocoronal emission lines such as 
Ly$\alpha$ and OI $\lambda1304$; these emission lines fill the spectrograph aperture and produce much broader line widths than their intrinsic widths (Eracleous \& Horne 1996). 
The feature in Figure 1c might be a resonant Mg or MgII line, though the scale height of Mg atoms is roughly $100\,\rm{km}$ lower than the {\it HST} altitude.
We note that Lyons et al. (1993b) reported a similar ``erratic" behavior of UV lines in FOS sky spectra, including the one at 2802 \AA. 
They found that these UV lines occur exclusively during daytime exposures, while the lines are absent on most of the daytime spectra taken.
Thus, we conclude that there might be a possible contribution from geocoronal emission lines to a daytime sky spectrum.

\subsection{Nighttime Data}

In order to avoid
possible contamination of the earthshine, we use only the nighttime data in the following analysis, although the contamination may be small in the twilight data as well
(see Figures 1 and 3 of Brown et al. 2000). 
When viewed from the Earth, the angle between the {\it HST} and the sun changes rapidly 
as the telescope orbits, and the orbital day/night 
status could switch from night to day or vice versa during an exposure. 
We obtained a list of day/night and night/day passing times for our observations, with the aid of the Space Telescope Science Institute (STScI) 
Help Desk, who kindly extracted the necessary telemetry data. 
STSDAS {\it deaccum} task was used to unbundle individual exposures from the original ACCUMlated data files. 
Exposure lengths ranged from 120 to 150 seconds, during which the sun - {\it HST} - target angle could change as large as $10^\circ$. 
The coordinates of the telescope were computed by the STSDAS {\it hstpos} task.
Using this list of day/night transition times, we assigned daytime fraction (pure nighttime = 0, pure daytime = 1) to individual exposures, and extracted only 
the pure night exposures. 

We searched for possible contamination by discrete sources (mostly Galactic stars) in two ways, and removed those contaminated exposures from the subsequent analyses. 
The first is to examine cutouts of the Sloan Digital Sky Survey (SDSS) and/or Digital Sky Survey images, 
using the Infrared Science Archive service at the Infrared Processing and Analysis Center. 
This process rejected discrete sources brighter than $22.5$ or $20.8$ Vega-magnitudes in the {\it B} or {\it R} band (Mickaelian 2004). 
The second is to look at the zeroth order spectra of the grating data. 
If a small discrete source affects the sky spectrum, its zeroth order spectral feature should be sharp, because such an object does not fill the aperture uniformly. 
In total, we identified 7 fields with visible discrete sources, and 4 fields with sharp zeroth order spectra.
Finally, we inspected all the individual spectra by eye, and removed those with apparent noise spikes. 
728 exposures in 54 fields survived the above cleaning process, whose total exposure time amounts to $93.0\,\rm{ksec}$; these data are summarized in Table 1.


Figure 2 displays the sky distribution of the above 54 fields, using the Mollweide projection in Galactic coordinates. 
Figure 3 presents 
the number of exposures as a function of (a) the Galactic latitude, (b) the ZL-subtracted $100\,\rm{\mu m}$ intensity 
(Schlegel et al. 1998), (c) the Galactic H$\alpha$ intensity (Finkbeiner 2003), and (d) the ZL intensity at $1.25\,\rm{\mu m}$ based on the DIRBE ZL model (Kelsall et al. 1998).

\begin{figure}
\begin{center}
 \includegraphics[scale=0.45]{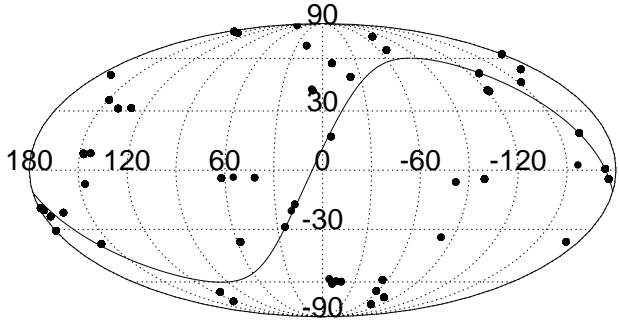} 
\end{center}
 \caption
 {Sky distribution of the FOS orbital nighttime data used in the present analysis, plotted with the Mollweide projection in Galactic coordinates.
 The solid line represents the ecliptic plane.}
\end{figure}

\begin{figure}
\begin{center}
 \includegraphics[scale=0.47]{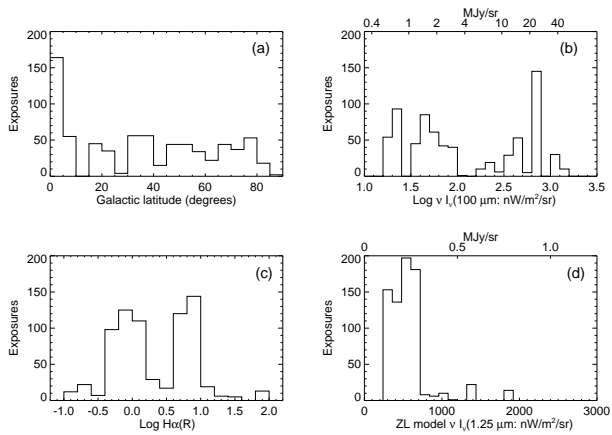} 
\end{center}
 \caption
 {Number of exposures as a function of (a) the Galactic latitude, (b) the ZL-subtracted $100\,\rm{\mu m}$ intensity, 
 (Schlegel et al. 1998), (c) the Galactic H$\alpha$ intensity (Finkbeiner 2003), and (d) the ZL intensity at $1.25\,\rm{\mu m}$ based on the DIRBE ZL model (Kelsall et al. 1998). 
}
\end{figure}

\begin{figure}
\begin{center}
 \includegraphics[scale=0.47]{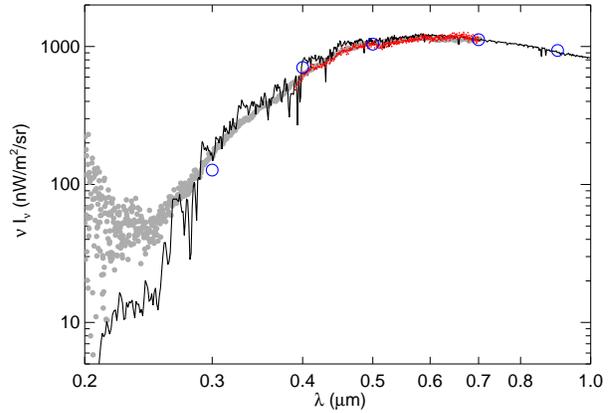} 
\end{center}
 \caption
 {FOS sky spectrum derived by integrating all the nighttime spectra regardless of the fields. 
The small and gray large dots represent the G650L and PRISM data, respectively. 
The solid line and the open circles represent the solar spectrum given by Colina, Bohlin, and Castelli (1996) 
and the ZL brightness presented by Leinert et al. (1998), respectively; they were arbitrarily scaled for ease of comparing the spectral shapes.
The FOS spectrum is much smoother than the solar spectrum because of the lower disperser resolution.
}
\end{figure}

\section{ANALYSIS}

Because the sky brightness is very low in the individual spectral elements, we perform the following analysis with the synthetic photometry in eight bands
from $0.2$ to $0.7\,\rm{\mu m}$, as summarized in Table 2.
We first checked the consistency of the flux calibration between the G650L and PRISM measurements in the $0.42$, $0.47$, $0.55$, and $0.65\,\rm{\mu m}$ bands, 
whether the two measurements are available.
As a result, we found a significant conflict between the G650L and PRISM brightness, whose origin is unknown at the moment.
Thus we decided to recalibrate the G650L measurements as follows, by referring to the earlier independent measurements.

The FOS twilight/nighttime spectrum of the parallel HDF field, which we discussed in Section 3.2, has the mean $0.55\,\rm{\mu m}$ brightness of $416\,\rm{nW\,m^{-2}\,sr^{-1}}$ 
(see Figure 1b).
On the other hand, 
Leinert et al. (1998; their table 16) presents the ZL brightness of $407\,\rm{nW\,m^{-2}\,sr^{-1}}$ at $0.5\,\rm{\mu m}$ in the same field at the time of our PRISM observations,
based on the ground-based measurements by Levasseur-Regourd and Dumont (1980) with a slight update.
These two independent measurements agree with each other within $2\%$.
Therefore,
we concluded that the PRISM/RED measurements are robust.
However, the G650 intensity is larger than the PRISM intensity by an almost uniform factor, $1.460 - 1.565$, 
in the four overlapping bands. 
Although the reason is not clear, this difference might be caused by flat-field calibration.
We thus assume that the PRISM intensity is correct, and scale down the G650L intensity by a uniform factor of 1.50.

Figure 4 presents the FOS sky spectrum compared with the solar spectrum (Colina et al. 1996) and the ZL spectrum (Leinert et al. 1998).
The sky spectrum was obtained by integrating all the exposures, regardless of the field coordinates. 
These spectra are similar to each other in the optical wavelength, suggesting that the ZL dominates the sky emission.
On the other hand,
the UV sky spectrum shows a significant excess over the solar spectrum. 
This may be due to strong time variation of the solar spectral irradiance in the UV, as reported by Ermolli et al. (2013) from the  SOlar Radiation and Climate Experiment.

\subsection{Models of the Emission Components}

Here we decompose the observed intensity of the diffuse sky emission into the three components, i.e., the ZL, the DGL, and the residual emission, in each of the eight 
photometric bands.
The EBL is included in the residual emission. 
The model brightness, $I_{\nu, i} ({\rm Model})$ at the $i$-th band, is defined as
\begin{equation}
I_{\nu, i} ({\rm Model}) = I_{\nu, i} ({\rm ZL}) + I_{\nu, i} ({\rm DGL}) + I_{\nu, i} ({\rm RSD}),
\end{equation}
where $I_{\nu, i} ({\rm ZL})$, $I_{\nu, i} ({\rm DGL})$, and $I_{\nu, i} ({\rm RSD})$ are the brightness of the ZL, DGL, and residual emission, respectively.

We minimize the following $\chi^2$ function:
\begin{eqnarray}
\chi_i^2 &=& \sum_j^{N_{{\rm exp}}^i} [I_{\nu, i} ({\rm Obs}) - I_{\nu, i} ({\rm Model})]^2/\sigma_{\nu, i}^2 \\
&=& \sum_j^{N_{\rm exp}^i} [I_{\nu, i} ({\rm Obs}) - I_{\nu, i} ({\rm ZL}) - I_{\nu, i} ({\rm DGL}) - I_{\nu, i} ({\rm RSD})]^2/\sigma_{\nu, i}^2, \\
\sigma^2_{\nu, i} &=& \sigma^2_{\nu, i} ({\rm Obs}) + \sigma^2_{\nu, i} ({\rm ZL}) + \sigma^2_{\nu, i} ({\rm DGL}), 
\end{eqnarray}
where $I_{\nu, i} ({\rm Obs})$ is the observed sky brightness, $j$ refers to the $j$-th exposure, and $N_{{\rm exp}}^i$ is the total number of exposures in the $i$-th band.
The quantities $\sigma_{\nu, i} ({\rm Obs})$, $\sigma_{\nu, i} ({\rm ZL})$, and $\sigma_{\nu, i} ({\rm DGL})$ are the uncertainty in the observed sky brightness, the ZL model, and the DGL model, respectively, and $\sigma_{\nu, i}$ represents their sum.
We describe the models of the ZL and DGL and the related uncertainties below.

\subsubsection{Zodiacal Light}

Because there are no ZL model from the UV to optical wavelengths, we estimate the ZL brightness by extrapolating the near-IR DIRBE ZL model to the shorter wavelength:
\begin{eqnarray}
I_{\nu, i} ({\rm ZL}) &=& a_i I_{\nu, i} ({\rm Sun}) D_{1.25}, \\
\sigma_{\nu, i} ({\rm ZL}) &=& 0.02 I_{\nu, i} ({\rm ZL}), 
\end{eqnarray}
where $a_i$ is the reflectance in the $i$-th band, 
and $D_{1.25}$ refers to the brightness of the DIRBE ZL model at $1.25\,\rm{\mu m}$, in units of $\rm{MJy\,sr^{-1}}$.
$I_{\nu, i} ({\rm Sun})$ is the $i$-th band intensity of the solar spectrum scaled to $1\,\rm{MJy\,sr^{-1}}$ at $1.25\,\rm{\mu m}$. 
It is derived from the solar spectrum given by Colina, Bohlin, and Castelli (1996). 
If the ZL spectrum is identical to the solar spectrum, then $a_i$ should be unity in all the eight bands. 
Thus we call $a_i$ the scaled ZL reflectance, 
which is unity at $1.25\,\rm{\mu m}$ by definition.
The factor of 0.02 in Equation (6) comes from the uncertainty in the DIRBE ZL model (Kelsall et al. 1998); 
here we assume that this value gives the statistical uncertainty of the ZL.

In the present analysis, 
the ZL spectrum is assumed to be isotropic throughout the sky.
Indeed, Tsumura et al. (2010) found such an isotropy of the ZL brightness at $0.8$--$1.25\,\rm{\mu m}$ from the CIBER/LRS measurements.
The ZL is similarly dominated by the scattered sunlight in both the FOS and CIBER/LRS spectral coverages.
However, the above ZL model may be too simplistic to fully account for the wavelength-dependent scattering phase function of the IPD grains, and may need further improvement in future works.

\subsubsection{Diffuse Galactic Light}

The DGL model brightness is defined as
\begin{eqnarray}
I_{\nu, i} ({\rm DGL}) &=& b_i I_{\nu, 100} - c_i I^2_{\nu, 100}, \\
I_{\nu, 100} &=& I_{\nu, {\rm SFD}} - 0.8\,\rm{MJy\,sr^{-1}}, \\
\sigma_{\nu, i}^2 ({\rm DGL}) &=& [b_i-2c_i I_{\nu, 100}]^2 \sigma^2_{\nu, 100}, 
\end{eqnarray}
where $b_i$ and $c_i$ are free parameters.
$I_{\nu, 100}$ is the $100\,\rm{\mu m}$ intensity 
from the ISM. 
$I_{\nu, {\rm SFD}}$ is the $100\,\rm{\mu m}$ intensity taken from the diffuse emission map 
of Schlegel et al. (1998; SFD hereafter). 

While the SFD $100\,\rm{\mu m}$ map has been processed to remove the ZL foreground, it may be contributed by emission components not associated with the ISM 
(hereafter non-ISM $100\,\rm{\mu m}$ emission), e.g., the EBL at this wavelength.
Equation (8) accounts for this non-ISM $100\,\rm{\mu m}$ emission, which is assumed to be $0.8\,\rm{MJy\,sr^{-1}}$.
Lagache et al. (2000) estimated its 
intensity 
of $0.78\pm0.21\,\rm{MJy\,sr^{-1}}$, using the data collected with the DIRBE and Far Infrared Absolute Spectrophotometer on board the {\it COBE}.
Matsuura et al. (2011) derived a similar non-ISM intensity of $0.67\pm0.19\,\rm{MJy\,sr^{-1}}$ at $90\,\rm{\mu m}$, from observations with 
the far-infrared surveyor on board the {\it AKARI} satellite.
The presence of non-ISM emission at this level ($\sim 0.8\,\rm{MJy\,sr^{-1}}$) was also supported by Matsuoka et al. (2011), in their analysis of the correlation
between the ZL-subtracted diffuse optical light and the SFD $100\,\rm{\mu m}$ map.
Planck Collaboration XXX (2014) estimated the non-ISM $100\,\rm{\mu m}$ emission to be $0.44\pm0.03\,\rm{MJy\,sr^{-1}}$ based on their extended halo model. 
We note that, in principle, this non-ISM $100\,\rm{\mu m}$ emission can include the (residual) ZL, EBL, and any other unknown components. 
For example, Dole et al. (2006) suggested that the non-ISM emission found by Lagache et al. (2000) includes the residual ZL of $0.3\,\rm{MJy\,sr^{-1}}$.

In the optically thin case, Equation (7) should be expressed as a linear correlation, $I_{\nu, i} ({\rm DGL}) = b_iI_{\nu, 100}$. 
Thus we call $b_i$ the slope coefficient or correlation slope, which is also presented as $\nu b_i = [3000/\lambda(\rm{\mu m})]$$b_i$ in units of ${\rm nW\,m^{-2}\,sr^{-1}/MJy\,sr^{-1}}$ below.
However, our analysis needs to deal with optically thick regions with up to $I_{\nu, 100} = 50\,\rm{MJy\,sr^{-1}}$, corresponding to the visual extinction $A_V \sim 5$ if we adopt $I_{\nu, 100} / A_V \sim 8 - 15\,\rm{MJy\,sr^{-1}\,mag^{-1}}$ 
(Ienaka et al. 2013). 
With a compilation of the 
optical data of high-latitude clouds, Ienaka et al. (2013) found that the correlation deviates from a linear function, i.e., 
the DGL intensity apparently saturates toward optically thick regions with high $100\,\rm{\mu m}$ intensity. 
They suggested that a negative $I^2_{\nu, 100}$ term should be introduced to the DGL models (Equation 7) to account for this saturation effect. 
At far-UV wavelengths where the optical depth is much larger than in the optical, the correlation becomes flatter 
with a larger scatter toward optically thick regions with $I_{\nu, 100} > 5\,\rm{MJy\,sr^{-1}}$ (Haikala et al. 1995; Sujatha et al. 2009; Murthy et al. 2010; Sujatha et al. 2010). 
This large scatter may be explained by not only the saturation, but also 
the variations in the intensity and spectral shape of the interstellar radiation field (ISRF); the ISRF can be quite different from that in optically thin regions, 
because optically thick regions may host star formation and hence many hot young stars. 

Our DGL model with a quadratic polynomial function 
might be too simple to fully account for the effects of the saturation and ISRF variation.
We checked this with an alternative model, which includes an additional cubic $I^3_{\nu, 100}$ term, 
and found that both the quadratic and cubic function models fit reasonably well to the data, i.e., 
there is no advantage of adding the cubic term.
We thus conclude that the quadratic function model in Equation (7) gives an adequate representation of 
the DGL for the present purpose. 

\subsubsection{Fitting Results}

\begin{figure}
\begin{center}
 \includegraphics[scale=0.45]{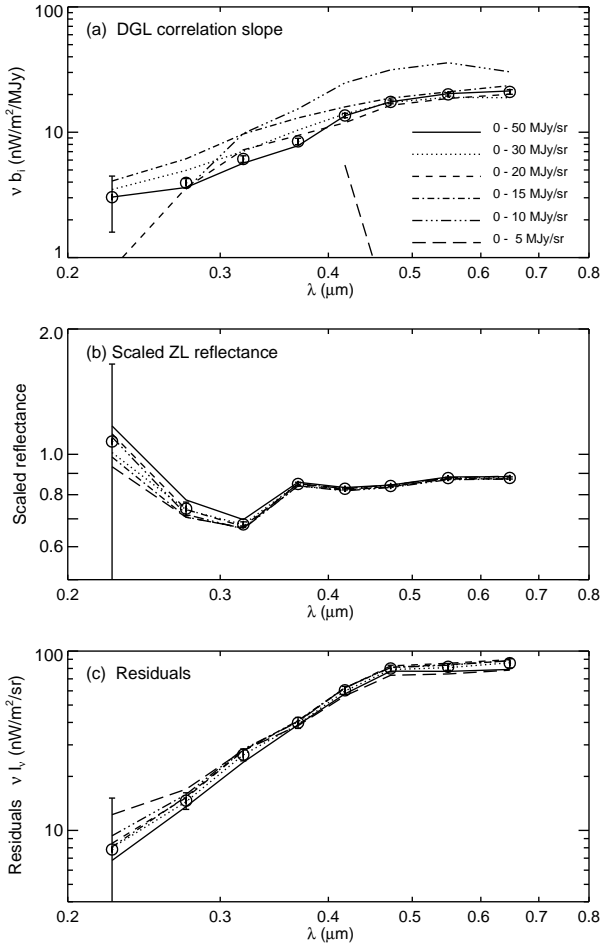} 
\end{center}
 \caption
 {
Fitting results of the six samples ($I_{\nu, 100} <$ 5, 10, 15, 20, 30, and $50\,\rm{MJy\,sr^{-1}}$), represented by the different types of lines as indicated in the lower right
of panel (a). 
The open circles represent the inverse-variance weighted mean of the four samples, $I_{\nu, 100} <$ 15, 20, 30, and $50\,\rm{MJy\,sr^{-1}}$ (see text). 
Panels (a), (b), and (c) present the DGL correlation slope $\nu b_i = [3000/\lambda(\rm{\mu m})]$$b_i$, the scaled ZL reflectance $a_i$, and the residual emission brightness
$\nu I_{\nu, i} ({\rm RSD})$, respectively. 
}
\end{figure}

We perform our fitting analysis in the six samples with the different maximum $100\,\rm{\mu m}$ intensity, namely, $I_{\nu, 100} <$ 5, 10, 15, 20, 30, and $50\,\rm{MJy\,sr^{-1}}$,
in order to examine the effect of DGL saturation in optically thick regions (see above).
The results are presented in Figure 5.
The DGL correlation slope $\nu b_i$ becomes negative in some photometric bands in the two samples, $I_{\nu, 100} <$ 5 and $10\,\rm{MJy\,sr^{-1}}$ (the negative points are represented 
by the missing parts of the curves in Figure 5a), which suggests that our DGL model failed to fit the data. 
This is likely 
caused by the 
limited fitting range of $I_{\nu, 100}$ in these smallest samples. 
The other fitting parameters 
were determined reasonably well in all the samples. 
In the following discussion, we use the inverse-variance weighted mean of the best-fit parameter values of the 
four samples, $I_{\nu, 100} <$ 15, 20, 30, and $50\,\rm{MJy\,sr^{-1}}$, 
as reported in Figure 5 and Table 2.
The quoted uncertainties are twice the error values obtained by a simple propagation of the errors of the individual samples, which takes into account the fact that the
four samples are not independent of each other.

\subsubsection{Contribution from Other Emission Components}

\begin{figure}
\begin{center}
 \includegraphics[scale=0.45]{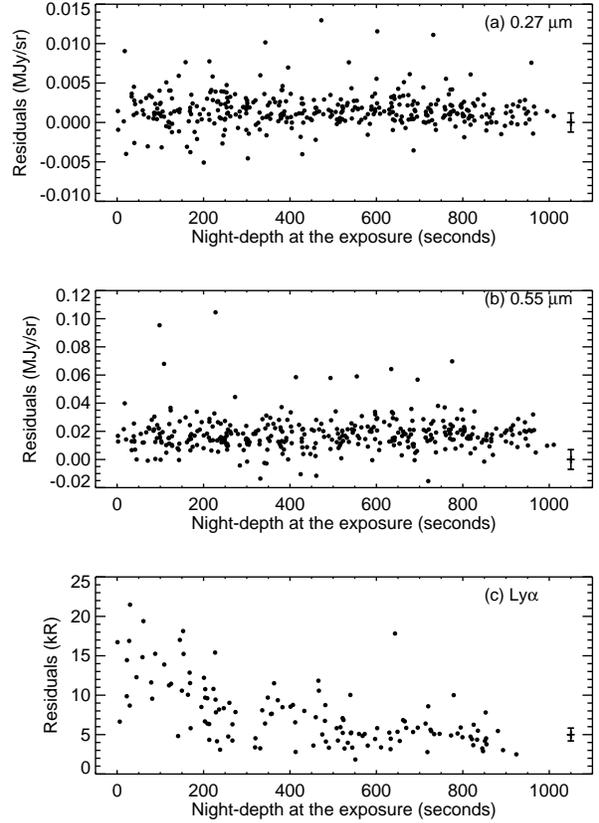} 
\end{center}
 \caption
 {Residual emission brightness $I_{\nu, i} ({\rm RSD})$ 
 as a function of the night-depth at $0.27\,\rm{\mu m}$ (panel a), $0.55\,\rm{\mu m}$ (panel b), and for Ly$\alpha$ (panel c).
 The typical error bars are presented in the lower right of each panel.
We use the full sample with $I_{\nu, 100} < 50\,\rm{MJy\,sr^{-1}}$ in this figure.
}
\end{figure}

Figures 6a and 6b present the brightness of the derived residual emission at $0.27$ and $0.55\,\rm{\mu m}$, as a function of the night depth.
The night depth is the time interval between day-to-night transition and the beginning of the exposure, or that between the end of the exposure and night-to-day transition, whichever is shorter.
The night-time duration is typically $\sim 2,000$ seconds, so the night depth of $\sim 1,000$ seconds corresponds to midnight. 
The figures exhibit no correlation, 
which confirms 
that the earthshine 
makes little contribution to our analysis. 

We further examined dependency of the residual emission brightness 
on various quantities, which include
the ZL model brightness, $100\,\rm{\mu m}$ intensity, Galactic latitude and longitude, zenith angle, limb angle, moon phase, and the solar MgII index that is related to the solar activity. 
No correlation is found, 
which indicates that our residual emission $I_{\nu, i} ({\rm RSD})$ is isotropic.
In particular, no correlation with the Galactic latitude suggests that Galactic stars are successfully removed 
(see Section 3.3) and that the residual emission contains little contribution from their radiation. 

The FOS SV observations used in this work also measured Ly$\alpha$ emission, using the BLUE digicon and the G150L grating. 
It provides an useful measure to examine the earthshine contribution to far-UV diffuse sky brightness, because the solar Ly$\alpha$ can diffuse into the night sky 
through resonant scattering by hydrogen atoms in the exosphere of the Earth (Eracleous \& Horne 1996). 
We decomposed the observed Ly$\alpha$ intensity into the ZL, DGL and residual emission in the same way as above, 
and found that the Ly$\alpha$ intensity is dominated by the residual emission, i.e., it has little correlation with the ZL or the SFD $100\,\rm{\mu m}$ brightness.
Figure 6c presents the derived Ly$\alpha$ residual brightness as a function of the night-depth.
The Ly$\alpha$ brightness decreases toward midnight, which suggests that the solar Ly$\alpha$ emission indeed diffuses into the night sky as described above. 
Similar correlations have been found in the far-UV count rates at the $0.153$ and $0.231\rm{\mu m}$ bands from the {\it Galaxy Evolution Explorer} ({\it GALEX}) observations (Sujatha et al 2010.). 

Figure 7 displays the residual emission brightness as a function of the Galactic H$\alpha$ intensity, taken from Finkbeiner (2003). 
No clear correlation is found, which is consistent with the idea that the Ly$\alpha$ emission is of geocoronal origin.

\begin{figure}
\begin{center}
 \includegraphics[scale=0.45]{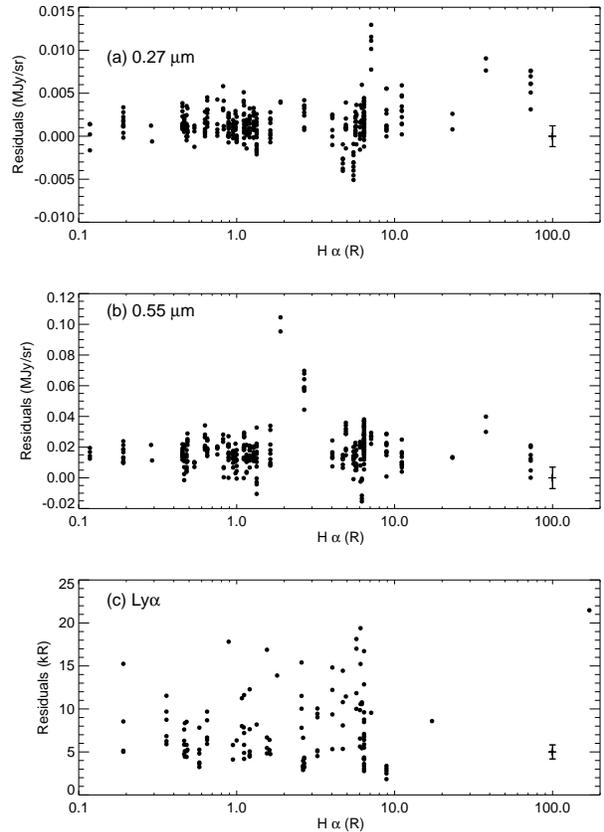} 
\end{center}
 \caption
 {Residual emission brightness $I_{\nu, i} ({\rm RSD})$ as a function of the Galactic H$\alpha$ intensity, at $0.27\,\rm{\mu m}$ (panel a), $0.55\,\rm{\mu m}$ (panel b), and for Ly$\alpha$ (panel c).
The typical error bars are presented in the lower right of each panel.
We use the full sample with $I_{\nu, 100} < 50\,\rm{MJy\,sr^{-1}}$ in this figure.
}
\end{figure}

\section{ZODIACAL LIGHT}

\begin{figure}
\begin{center}
 \includegraphics[scale=0.53]{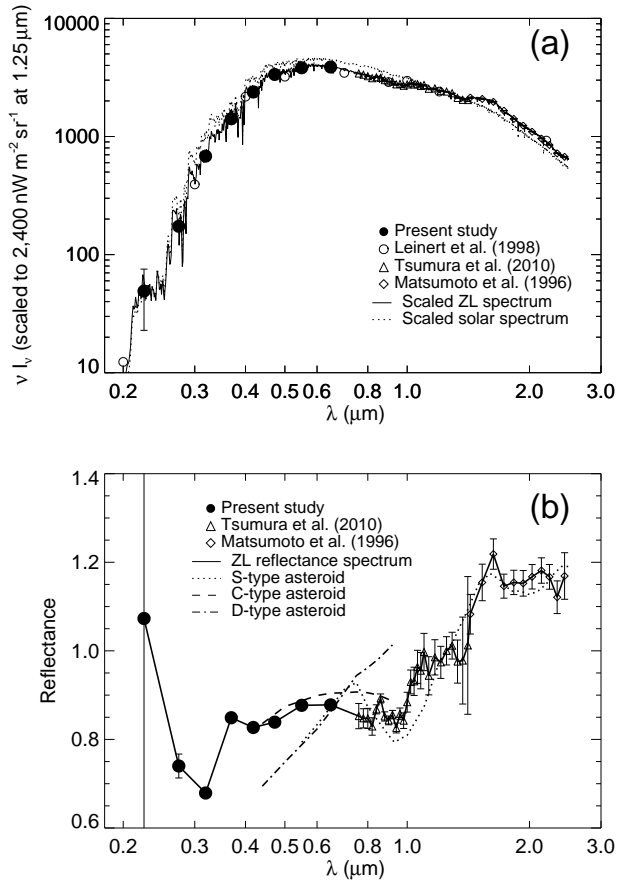} 
\end{center}
 \caption
 {Panel (a): 
The ZL brightness derived in this work (filled circles) and those measured by Tsumura et al. (2010; triangles) and Matsumoto et al. (1996; diamonds).
The ZL and solar spectra are represented by the solid and dashed lines, respectively.
The open circles show the ZL brightness toward ($\lambda - \lambda_{\odot} = 90^\circ, \beta = 0^\circ$), presented by Leinert et al. (1998). 
All the data are scaled to $1\,\rm{MJy\,sr^{-1}} = 2,400\,{\rm nW\,m^{-2}\,sr^{-1}}$ at $1.25\,\rm{\mu m}$.
Panel (b): 
The ZL reflectance derived in this work (filled circles) and those measured by Tsumura et al. (2010; triangles) and Matsumoto et al. (1996; diamonds).
The ZL reflectance spectrum, which connects the above data points, is represented by the solid line.
These spectra are normalized at $1.25\,\rm{\mu m}$.
The arbitrarily scaled reflectance spectrum of an S-type asteroid 25143-Itokawa (Binzel et al. 2001), typical C-type, and D-type (cometary) asteroids (Bus \& Binzel 2002)
are represented by the dotted, dashed, and dot-dashed line, respectively. 
}
\end{figure}

Figure 8a presents the derived ZL spectrum, which is a product of the solar spectrum and the ZL reflectance $a_i$ (linearly interpolated in wavelength).
All the spectra in this panel are scaled to $1\,\rm{MJy\,sr^{-1}} = 2,400\,{\rm nW\,m^{-2}\,sr^{-1}}$ at $1.25\,\rm{\mu m}$, for ease of mutual comparison.
Our result is in good agreement with the ZL spectrum presented by Leinert et al. (1998, toward $\lambda - \lambda_{\odot} = 90^\circ, \beta = 0^\circ$).
We also show the IR results taken from the CIBER/LRS measurements (Tsumura et al. 2010) and the {\it IRTS}/NIRS measurements (Matsumoto et al. 1996).
The ZL spectrum is redder than the solar spectrum at $> 1.5\,\rm{\mu m}$, which may point to major contribution of large IPD particles ($>1\,\rm{\mu m}$) to the near-IR ZL (Matsuura et al. 1995).

Figure 8b presents the derived ZL reflectance, along with the IR results presented by Tsumura et al. (2010) and Matsumoto et al. (1996).
In the near-IR, 
Tsumura et al. (2010) suggested that the ZL is dominated by S-type asteroidal dust,
based on the similarity of their reflectance spectra at $> 1.5\,\rm{\mu m}$.
On the other hand, we found that the reflectance spectrum becomes much flatter in the optical. 
This indicates a large contribution to the ZL dust from C-type asteroids, which have a similarly flat reflectance spectrum in this wavelength range. 
A similar trend was reported by Lumme and Bowell (1985), who suggested that the ZL color closely resembles the colors of C-type asteroids.
In addition, some CM chondrites reportedly have flat reflectance spectra in the optical (Vernazza et al. 2015).
These results may also support the C-type asteroid origin of the IPD particles.

Yang and Ishiguro (2015) suggested that the most of the IPD particles originate from comets (D-type asteroids), by comparing optical properties (i.e., albedo and spectral gradient) of various asteroids (Bus \& Binzel 2002) with those of the ZL (Ishiguro et al. 2013).
This result is consistent with a numerical simulation, which takes into account kinematic and dynamical processes of the IPD (Nesvorn\'y et al. 2010).
However, we found that the ZL spectrum differs significantly from that of D-type asteroids (see Figure 8b), which is in conflict with
the cometary origin of the IPD responsible for the optical ZL.

It is worth noting 
a dip in the ZL reflectance spectrum seen at around $0.3\,\rm{\mu m}$. 
Such a UV absorption feature has been reported in earlier studies,
which measured reflectance spectra in various samples of asteroids and meteorites in laboratory experiments (e.g., Hiroi et al. 1996; Yamada et al. 1999; Matsuoka et al. 2015).
Dulay and Lazarev (2004) proposed the $\pi-\pi^*$ prasmon resonance in polycyclic aromatic hydrocarbon molecules as an origin of this absorption, 
while Cloutis et al. (2008) raised a possibility of Fe-O charge transfer transition in minerals, 
which is expected to produce 
several absorption features at $0.2 - 0.4\,\rm{\mu m}$.
The exact nature of this spectral dip will be identified in future ZL measurements with much higher spectral resolution than available now.

\section{DIFFUSE GALACTIC LIGHT}

\begin{figure}
\begin{center}
 \includegraphics[scale=0.53]{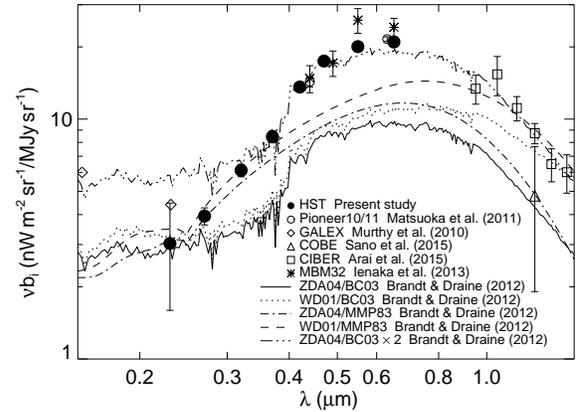} 
\end{center}
 \caption
 {
The DGL correlation slope $\nu b_i$ as a function of wavelength, compiled from this work (filled circles), 
the {\it Pioneer 10/11} measurements (Matsuoka et al. 2011; open circles), 
the {\it GALEX} measurements (Murthy et al. 2010; open diamonds), 
the CIBER measurements (Arai et al. 2015; open squares), 
the DIRBE measurements (Sano et al. 2015; open triangle), and the ground-based measurements toward a high Galactic latitude translucent cloud (Ienaka et al. 2013; asterisks).
The synthetic DGL spectra of Brandt and Draine (2012), toward a high Galactic latitude ($b = 60^{\circ}$), are displayed with the four different lines: 
ZDA04/BC03 (solid), WD01/BC03 (dotted), ZDA04/MMP83 (dot-dashed), and WD01/MMP83 (dashed). 
The dot-dot-dashed line represents the ZDA04/BC03 model scaled to the observed values in the optical.}
\end{figure}

Figure 9 presents 
the derived DGL correlation slope $\nu b_i$,  along with the results from earlier measurements.
As we discussed in Section 4.1.2, 
the correlation slope $\nu b_i$ of our model (see Equation 7) provides 
the DGL spectrum in optically thin regions. 
A clear $4000\,{\rm \AA}$ break is observed in the DGL 
spectrum, which was also found by Brandt and Draine (2012) in their analysis of 
SDSS sky spectra.

Brandt and Draine (2012) studied the relation between the dust-scattered light and the $100\,\rm{\mu m}$ emission in optically thin ISM,
with single-scattering radiative transfer calculations assuming a plane parallel galaxy. 
They 
used two models 
of the local ISRF continua based on Mathis, Mezger, and Panagia (1983; MMP83) and Bruzual and Charlot (2003; BC03), combined with two dust models based on
Zubko, Dwek, and Arendt (2004; ZDA04) and Weingartner and Draine (2001; WD01). 
The MMP83 model is a modified version of the original MMP83 model, consisting of four dilute blackbodies to which the ISRF dereddening has been applied (see Brandt \& Draine 2012, for details).
The BC03 model is a stellar population synthesis model, 
assuming solar metallicity and an exponentially-declining star formation rate of $\propto \exp(-t/5\, {\rm Gyr})$.
The WD01 model contains larger dust grains than in the ZDA04, and produces a redder scattered spectrum in the optical and near-IR wavelengths.

Figure 9 indicates that the predictions of the Brandt and Draine (2012) models 
underestimate the observed DGL correlation slopes by a factor of two.
The models can reproduce the observations only when an arbitrary scaling is allowed, as demonstrated by the dot-dot-dashed line in the figure.
The reason for this discrepancy may be attributed to the effect of anisotropic scattering of the dust grains.
Due to the forward scattering, the intensity ratio of the scattered light to $100\,\rm{\mu m}$ emission is expected to increase toward low Galactic latitude (Jura 1979).
In fact, Sano et al. (2016b) analyzed the DIRBE data at $1.25$ and $2.2\,\rm{\mu m}$  and found that the above ratio increases toward low Galactic latitude by a factor of two.
While the DGL models of Brandt and Draine (2012) are only valid in high Galactic latitudes, the present analysis includes the observed fields distributed 
throughout the sky, including the low Galactic latitudes.
This may be the reason for the higher intensity ratio we found compared to the model predictions.

\section{RESIDUAL EMISSION}

\begin{table*}
 \renewcommand{\arraystretch}{1.0}
 \caption{Decomposition results of the FOS sky spectra}
\begin{center}
  \label{symbols}
  \scalebox{0.8}{
  \begin{tabular}{lcccccccc}
  \hline
   Band & $0.23\,\rm{\mu m}$ & $0.27\,\rm{\mu m}$ & $0.32\,\rm{\mu m}$ & $0.37\,\rm{\mu m}$ & $0.42\,\rm{\mu m}$ & $0.47\,\rm{\mu m}$ & $0.55\,\rm{\mu m}$ & $0.65\,\rm{\mu m}$ \\
   \hline
   Effective $\lambda$ ($\rm{\mu m}$) $^a$ & 0.225 & 0.274 & 0.319 & 0.369 & 0.418 & 0.472 & 0.550 & 0.648 \\
   $\lambda$ range ($\rm{\mu m}$) $^b$ & 0.20--0.24 & 0.24--0.29 & 0.29--0.34 & 0.34--0.39 & 0.39--0.44 & 0.44--0.50 & 0.50--0.60 & 0.60--0.70\\
   Disperser $^c$ & P & P & P & P & G \& P & G \& P & G \& P & G \& P \\
   \\
   \multicolumn{9}{l}{ZL reflectance (scaled to unity at $1.25\,\rm{\mu m}$)}\\
   $a_i$ & 1.1 & 0.74 & 0.68 & 0.849 & 0.827 & 0.839 & 0.877 & 0.878\\
   $\sigma_{\nu, i}(a_i)$ & 0.6 & 0.03 & 0.01 & 0.008 & 0.005 & 0.004 & 0.004 & 0.005\\
   \\   
   \multicolumn{9}{l}{DGL correlation slope (${\rm nW\,m^{-2}\,sr^{-1}/MJy\,sr^{-1}}$; $\nu b_i = [3000/\lambda (\rm{\mu m})]$$b_i$)} \\
   $\nu b_i$ & 3.0 & 3.9 & 6.1 & 8.5 & 13.6 & 17.5 & 20.1 & 21.0\\
   $\sigma_{\nu, i}(\nu b_i)$ & 1.4 & 0.3 & 0.4 & 0.5 & 0.5 & 0.6 & 0.6 & 0.9\\
   \\
   \multicolumn{9}{l}{DGL quadratic coefficient ($10^5\,{\rm (MJy\,sr^{-1})^{-1}}$; see Equation 7)}\\
   $c_i$ & 0.1 & 0.3 & 0.7 & 1.1 & 2.4 & 3.3 & 4.4 & 4.5\\
   $\sigma_{\nu, i}(c_i)$ & 0.4 & 0.1 & 0.1 & 0.2 & 0.3 & 0.3 & 0.4 & 0.7\\
   \\
   \multicolumn{9}{l}{Residual emission brightness (${\rm nW\,m^{-2}\,sr^{-1}}$)}\\
   $\nu I_{\nu,i} ({\rm RSD})$ & 7.8 & 14.7 & 26.4 & 39.8 & 60.4 & 80.0 & 81.7 & 85.6\\
   $\sigma[\nu I_{\nu,i} ({\rm RSD})]$  & 7.3 & 1.6 & 2.0 & 2.8 & 3.1 & 3.4 & 3.6 & 5.0\\
   $\sigma[\nu I_{\nu,i} ({\rm Sys})]$ $^d$ & 0.3 & 1.1 & 4.3 & 8.8 & 14.9 & 20.9 & 23.9 & 24.2\\
   $\sigma[\nu I_{\nu,i} ({\rm Cal})]$ $^e$ & 0.4 & 0.7 & 1.3 & 2.0 & 3.0 & 4.0 & 4.1 & 4.3\\
   $\sigma[\nu I_{\nu,i} ({\rm Tot})]$ $^f$ & 7.3  & 2.1 & 4.9 & 9.4 & 15.5 & 21.5 & 24.5 & 25.1\\
   \\
   \multicolumn{9}{l}{Solar spectrum in units of ${\rm nW\,m^{-2}\,sr^{-1}}$ (scaled to $2,400\,{\rm nW\,m^{-2}\,sr^{-1}} = 1\,{\rm MJy\,sr^{-1}}$ at $1.25\,\rm{\mu m}$)}\\
   $\nu I_{\nu,i} ({\rm Sun})$ & 45.8 & 234.6 & 1004.4 & 1665.0 & 2876.6 & 3993.4 & 4352.7 & 4411.6\\
   
    \hline
    \end{tabular}
    }
    \end{center}
    \medskip
   
    $^a$ --- Effective wavelength of each band.\\ 
    $^b$ --- Lower and upper cut-off wavelengths.\\ 
    $^c$ --- ``P'' and ``G'' indicate the PRISM and G650L, respectively.\\
    $^d$ --- Systematic uncertainty of the DIRBE ZL model. \\
    $^e$ --- Calibration uncertainty assumed to be $5\%$ of $\nu I_{\nu,i} ({\rm RSD})$.\\
    $^f$ --- Total uncertainty of the residual emission brightnesss. 
         
 \end{table*}

Before discussing the derived residual emission, 
we estimate its additional uncertainty, 
which comes from the DIRBE ZL model and the absolute flux calibration. 
Kelsall et al. (1998) estimated the systematic uncertainty of their ZL model to be $15\,{\rm nW\,m^{-2}\,sr^{-1}}$ at $1.25\,\rm{\mu m}$, from the difference between 
their two typical models.
The uncertainty in the absolute flux calibration 
is difficult to estimate in nature, which we
conservatively 
assume to be $5\%$ of the derived residual emission.
The total uncertainty, $\sigma[\nu I_{\nu,i} ({\rm Tot})]$, is calculated as the quadratic sum of all the error components, i.e.,  the statistical error $\sigma[\nu I_{\nu,i} ({\rm RSD})]$, 
the ZL model error $\sigma[\nu I_{\nu,i} ({\rm Sys})]$, and the flux calibration error $\sigma[\nu I_{\nu,i} ({\rm Cal})]$.
The individual error budget and their sum are listed in Table 2.
The systematic uncertainty of the ZL model is dominant over the statistical and calibration uncertainty in most of the cases.


Figure 10 compares the derived residual emission brightness 
with the earlier results 
and with the brightness of the IGL.
Our residual emission is 
several times brighter than the observed or modeled IGL (Xu et al. 2005; Gardner, Brown, and Ferguson 2000; Totani et al. 2001; Madau and Pozzetti 2000;
Fazio et al. 2004; Dom\`inguez et al. 2011), and also exceeds the EBL estimate
obtained with {\it Pioneer 10/11} (Matsuoka et al. 2011).
On the other hand, Bernstein (2007) reported a similarly bright residual emission to ours from their {\it HST}/WFPC2 measurements.

\begin{figure*}
\begin{center}
 \includegraphics[scale=0.9]{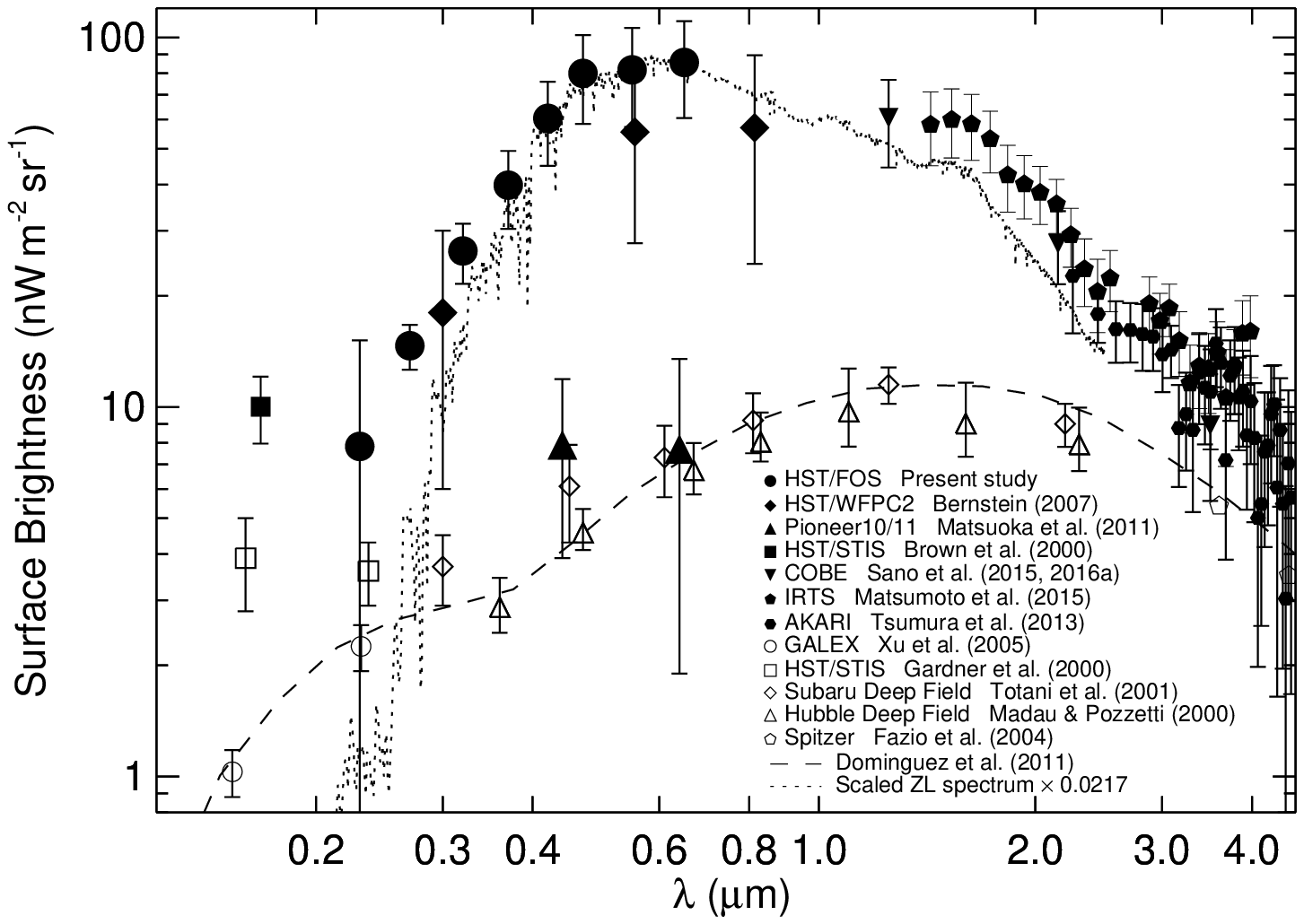} 
\end{center}
 \caption
 {Compilation of the residual emission brightness (filled symbols) and the IGL brightness (open symbols and the dashed line) from the UV to near-IR wavelengths.
The present results are indicated by the filled circles; the references for the other symbols are described at the bottom right corner.
The dotted line represents an arbitrarily-scaled ZL spectrum. 
}
\end{figure*}

Figure 10 demonstrates that the spectrum of the residual emission is strikingly similar to that of the ZL. 
Such a similarly has already been pointed out in the near-IR by Dwek, Arendt, and Krennrich (2005).
It may indicate the existence of an additional component to the known, currently-modeled ZL.
If we assume that the difference between the present FOS residual and the {\it Pioneer 10/11} EBL estimate is all attributed to a missed ZL component, 
and that the known and missed ZL components have the same spectrum, then 
the resultant intensity of the missed ZL is ($0.0217\pm0.0007$) times that of the known ZL scaled to $1\,\rm{MJy\,sr^{-1}}$ at $1.25\,\rm{\mu m}$.
It corresponds to roughly $10\%$ of the known ZL intensity ($\sim0.2$--$0.3\,\rm{MJy\,sr^{-1}}$ at $1.25\,\rm{\mu m}$)
measured in the present data.

Presence of a missed ZL component is not a new idea.
The parameters of the analytical functions in the DIRBE ZL model have been determined by fitting 
the seasonal variation of the observed brightness toward a grid of directions over the sky. 
Hauser et al. (1998) emphasized that ``this method cannot uniquely determine the true ZL signal; in particular, an arbitrary isotropic component could be added to the model without affecting the parameter values determined in the fitting to the seasonal variation of the signal.''
Recently, by using the {\it AKARI} diffuse emission maps at $9$ and $18\,\rm{\mu m}$, Kondo et al. (2016) created a new zodiacal emission model that takes into account an isotropic emission component, and indeed found a signal of such a component
that probably originates from the IPD.

In the 
long term, observations of the diffuse sky emission from beyond the ZL cloud are required to understand the origin of the excess emission over the IGL.
Such observations will be carried out by, e.g.,  
the EXo-Zodiacal Infrared Telescope (EXZIT), which will be
 one of the instruments on board the Solar Power Sail mission to Jupiter in 2020s (Matsuura et al. 2014).
The EXZIT will 
allow us to measure the ZL-free diffuse sky, and thus draw a firm conclusion on the radiation coming truly from outside the solar system.

\section{SUMMARY AND CONCLUSION}

We analyzed the {\it HST}/FOS blank sky spectra from the UV to optical wavelengths (0.2 -- $0.7\,\rm{\mu m}$). 
The observations were performed toward 54 fields, which were distributed widely over the sky. 
We found that the daytime spectra are contaminated by the earthshine, and 
hence used the data taken only in 
the orbital nighttime.
We defined eight photometric bands spanning our spectral coverage, and decomposed the observed intensity into the ZL, DGL, and residual emission components
in each band.

We found that the ZL reflectance spectrum is relatively flat, except for a dip seen at around $0.3\,\rm{\mu m}$, which indicates major contribution of C-type asteroids to the IPD.
The intensity ratio of the DGL to $100\,\rm{\mu m}$ emission has a similar spectral shape to the model predictions (Brandt \& Draine 2012), but is larger than the model by a factor of two. 
Such discrepancy may be caused by the difference in Galactic latitudes between our observations and the model calculations.

The residual emission has sometimes been regarded as the EBL in previous studies. 
However, we found that the derived residual emission has a similar spectral shape to that of the ZL, and is much brighter 
than the EBL estimate from the {\it Pioneer10/11} measurements at beyond the IPD cloud.
Assuming that this excess emission is due to an unknown ZL component, 
we provided a quantitative estimate of its intensity.
Our analysis revealed a likely presence of a missed component in the current DIRBE ZL model, which manifests itself most clearly in the similar spectral shapes
of the residual emission and the ZL from the UV to near-IR wavelength, as one observes in Figure 10.

\begin{ack}
We are grateful to the referee, Masateru Ishiguro, for his very useful comments to improve this paper.
We would also like to thank the staff members of the STScI Help Desk and the engineers, for providing the day/night transition time data for 
the observations used in the present analysis;
this work would not have been completed without their dedication.
We thank Y. Yoshii, T. Totani, T. Yamamoto, T. Kozasa, T. Matsumoto, S. Matsuura, M. Matsuoka, and A. Takahashi for useful discussions and encouragement. 
This work has been supported by Grants-in-Aid for Specially Promoted Research on Innovative Areas (22111503, 24111705).
\end{ack}





\end{document}